\documentclass{article}
\usepackage{epsfig}
\begin{document}
\title{
Quantum phase transition in a random-tiling model}
\author{D. K. Sunko\thanks{email: dks@phy.hr.
Published in Fizika A (Zagreb) \textbf{8} (1999) 4, 311--318.}\\
Department of Physics,\\Faculty of Science,\\
University of Zagreb,\\
Bijeni\v cka cesta 32,\\
HR-10000 Zagreb, Croatia.}
\date{}

\maketitle

\noindent
{\small\bf Dedicated to Professor Boran Leonti\'c on the occassion of
his 70$^\mathbf{th}$ birthday}
\begin{abstract}
The analogue of a Mott-Hubbard transition is discussed, which appears at an
incommensurate filling in a model of a two-dimensional plane, randomly tiled
with CuO$_4$ `molecules', simulating the copper-oxide planes of high-T$_c$
superconductors. It is shown to be a quantum phase transition, which can be
crossed either in doping, at a fixed hopping overlap $t$, or in $t$, when
the doping is fixed in a certain range below half-filling. It is first-order,
closely analogous to a liquid-gas transition.
\end{abstract}

\section{Introduction}

Quantum phase transitions are sudden changes in the nature of the ground
state of a physical system, when some dynamical parameter reaches a critical
value. Their theoretical mark of distinction is that they occur even when the
temperature is strictly zero (as only theory can make it), so that the
sampling of phase space, responsible for finding the new optimal
configuration, is entirely by quantum, not thermal, fluctuations.

The main interest in these transitions is that they are triggered by quantum
many-body effects, so they indicate the appearance of new `states of matter'
built by microscopic interactions. The processes involved in them have been
well understood in many cases where a weak-coupling limit is appropriate, the
most famous of which is BCS superconductivity. In purely electronic systems
in the strong-coupling limit, to which the subsequent discussion is limited,
apart from numerical simulations~\cite{dagotto} and one-dimensional
systems~\cite{emery}, the study of phase transitions has most often resorted
to oblique approaches: drawing analogies with the one-dimensional
case~\cite{anderson}, making inferences from sophisticated weak-coupling
studies~\cite{dzaloshinski,rice}, or solving essentially one-body problems,
with some constraint added, which supposedly accounts for the effects of
strong correlations~\cite{kotliar,krauth}.

The present work is of this last kind. It is motivated by the strange `normal
state' of hole-doped high-temperature superconductors, which conduct
electricity in some way which has so far defied considerable efforts at
explanation. These have revolved around the question, whether the conducting
state can be understood by modifying a Fermi liquid picture~\cite{varma}, or
some radically different zeroth-order approximation, involving perhaps
spin-charge separation~\cite{anderson}, is needed. In the materials
themselves, a wide range of measurements, from conductivity~\cite{magn} to
photoemission~\cite{fujimori}, indicate a crossover, from a strange to an
apparently Fermi-liquid electron system, as the doping increases from
underdoped to overdoped.

Since the present model is one-body, it is unable to address these `deep'
issues directly, being, in addition, tied to a Fermi liquid language by a
formal construction. However, it can still distinguish between a liquid and a
gas. It turns out that the Mott-Hubbard transition in the model is a
liquid-to-gas transition in the direction of \emph{increasing} doping. The
system is a liquid in the lower Hubbard band, and a gas in the in-gap band.
This seems counterintuitive, because the in-gap band corresponds to a more
crowded real space. It is due to the transition being provoked by hopping
fluctuations, which find the spatially less ordered `gas' beneficial for
delocalization, precisely when crowding is high. The remainder of the article
is a brief elaboration of these points. In particular, it will be shown that
the transition is a quantum one, and can be triggered by increasing the
hopping overlap at zero temperature, when the doping is fixed in a certain
narrow range below half-filling.

\section{The range of the transition}

The model is the same as described previously~\cite{sunko}. It is a random
tiling of the copper-oxide planes by CuO$_4$ `molecules' consisting of a
copper site connected with the neighboring oxygens by a tight-binding overlap
$t$. The only other parameter is the copper-oxygen energy splitting,
$\Delta_{pd}>0$ in the hole picture. Only up-spins actually hop over these
molecules, while the presence of the down-spins is expressed by some given
concentration of forbidden sites, i.e. the absence of molecules over which
the up-spins can hop. Hubbard's repulsion $U$ is thus effectively infinite,
while hopping is `projected': the presence of a down-spin on a site cancels
the up-spin hopping to that site. The fact that down-electrons are really not
heavier than the others is simulated by annealing, so that the phase space
accessible to mobile electrons includes all possible positions of the static
ones, in contrast to the `quenched' approach, which would be appropriate for
real heavy impurities. The resulting translational invariance enables a
$k$-space formulation which respects the Pauli principle, so the model
effectively interpolates between quantum order in inverse space at low
temperature, and classical disorder in real space at high temperature. Which
temperature is `low' is determined by the width $W$ of the in-gap band,
created by the transition:
\begin{equation}
W={\widetilde{\Delta}_{pd}\over
1+\Delta_{pd}/(2\widetilde{\Delta}_{pd})},
\end{equation}
where $\widetilde{\Delta}_{pd}=\sqrt{\Delta_{pd}^2/4+4t^2}-\Delta_{pd}/2$ is
the distance from the oxygen band to the middle of the in-gap band.

\begin{figure}[t]
\begin{center}
\begin{picture}(11.52,7.2)
\put(0,0){\psfig{
figure=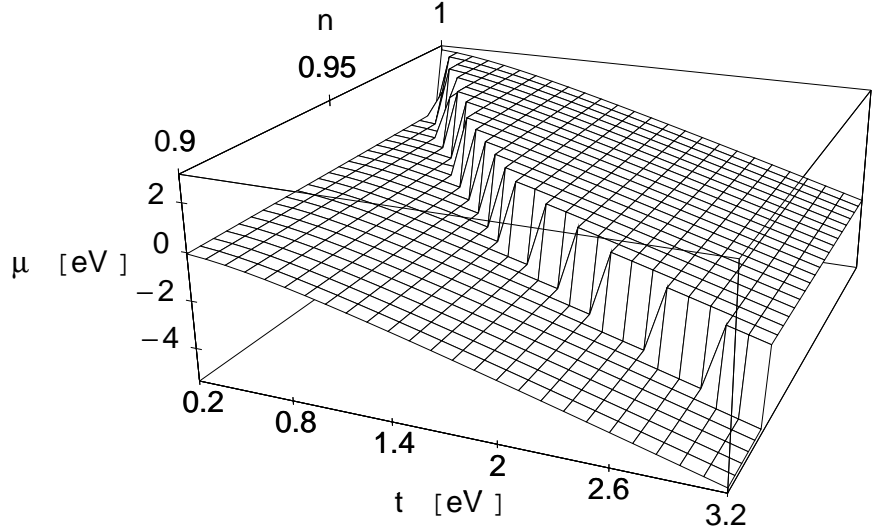
,width=11.52cm,height=7.2cm}}
\end{picture}
\end{center}
\caption{The chemical potential $\mu$ of the mobile spins, as a function of
the hopping overlap $t$ and doping $n=2n_\uparrow=2n_\downarrow$. Electron
doping is $n<1$, and $\Delta_{pd}=3$~eV, $T=40$~K throughout.
\label{prijelaz}}
\end{figure}
Figure \ref{prijelaz} shows the Mott-Hubbard transition in the model, as a
function of concentration $n$ and hopping overlap $t$. Clearly, the simple
`kinematical' expectation that it would occur at half-filling for all $t$ is
not fulfilled. Instead, the greater $t$ is, the sooner will the transition
occur in doping; however, there is also a saturation effect, so even a very
large $t$ will not pull the transition below $n\approx 0.9$. When
$t/\Delta_{pd}$ is small, it occurs near half-filling.

It can be shown that at zero temperature, the transition will occur at the
doping value
\begin{equation}
n_\uparrow+n_\downarrow=1+{\sin^2(\varphi/2)\over 2}
\int_{BZ}\left[\cos k_x+\cos k_y
\right]f_{\mu}(\varepsilon_-),
\label{crit}
\end{equation}
where $\sin\varphi=2t/\sqrt{\Delta_{pd}^2/4+4t^2}$, and the Fermi function is
in terms of the effective bonding band dispersion $\varepsilon_-$ of the
mobile, up-spins. (The chemical potential $\mu$ refers to them,
$n_\uparrow=\int_{BZ}f_{\mu}$.) Since the band is being filled from the edge
of the zone, the factor in brackets is negative, moving the transition from
the classical line $n_\uparrow+n_\downarrow=1$ to smaller values,
corresponding to electron doping. If $t$ were so large that $\sin\varphi\to
1$, the correction would be $-1/\pi^2\sim -0.1$ at half-filling, accounting
for the saturation in figure \ref{prijelaz}.

Equation (\ref{crit}) shows the transition to be controlled by the hopping
overlap $t$. It is a quantum phase transition, with (\ref{crit}) giving the
critical line in the plane of $t$ and doping, at zero temperature. The
physical origin of equation (\ref{crit}) is in a basic assumption of the
model, that the presence of forbidden sites influences the effective
dispersion of the mobile spins $\varepsilon_-$ through a bulk parameter, the
chemical potential $\nu$ of the static down-spins, schematically:
\begin{equation}
Z_{\mu,\nu}=Z_\mu[\varepsilon_-(\nu)]\cdot Z_\nu.
\label{fact}
\end{equation}
This is, essentially, a thermodynamical assumption: the energy levels of
mobile up-spins depend on $\nu$ as a `mechanical' measure of their available
phase space. Equation (\ref{fact}) implies a contribution from
$\partial\varepsilon_-(\nu)/\partial\nu$ to the counting of down-spins, which
then produces a transition in the middle of the band; though not precisely in
the middle, as shown above. The reason for this can be understood along the
same lines: the band \emph{narrows} as the chemical potential of down-spins
increases, so that $\partial\varepsilon_-/\partial\nu$ cannot be the same for
all states in the band. When it is integrated over the Brillouin zone, it
gives the correction in equation (\ref{crit}).

The point of all this is that as soon as one imagines something like
(\ref{fact}) is possible, the transition will occur at an incommensurate
filling; this is a very robust consequence of the assumption (\ref{fact}),
not the effect of some detail. As seen from equation (\ref{crit}), neglecting
the correction would be like taking $\cos k_x+\cos k_y=0$, true at
half-filling, to be true for all states. The present model, by contrast,
takes the Pauli principle for the mobile electrons into account exactly. The
very existence of a quantum dispersion pushes the transition away from
half-filling.

[To derive equation (\ref{crit}), one simply adds the saddle-point equations
\begin{equation}
n_\uparrow={\partial\ln Z_{\mu,\nu}\over\partial\beta\mu},\,
n_\downarrow={\partial\ln Z_{\mu,\nu}\over\partial\beta\nu},
\end{equation}
inserting at the same time the particular limit $\nu\to\varepsilon_p-0$,
which corresponds to their solution at the critical line and $T=0$. Here
$\varepsilon_p$ is the bare energy of the oxygen level.]

\section{The nature of the transition}\label{trans}

The model transition from the lower Hubbard to the in-gap band is like from a
liquid to a gas. It has been noticed previously~\cite{sunko} that the entropy
of mobile spins suffers a discontinuous jump at the transition, and that the
contribution of the interaction to the entropy abruptly rises from a fairly
large negative value ($\sim\,-0.2$~k$_B$ per site) to zero. These are
characteristics of a first-order liquid-gas transition. In words, as soon as
the interaction has created the in-gap band, its effects are absorbed into
the one-particle properties of the band states.

\begin{figure}[t]
\begin{center}
\begin{picture}(11.18,9.64)
\put(0,0){\psfig{
figure=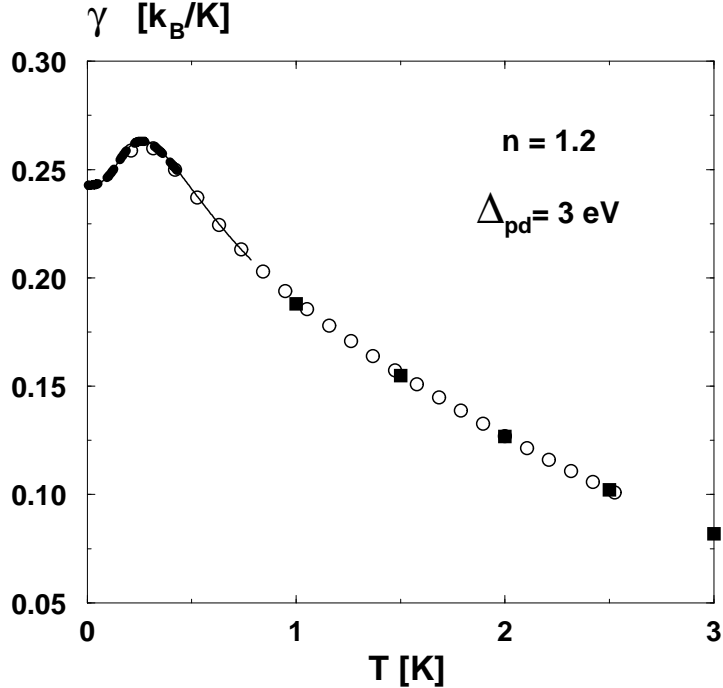
,width=11.18cm,height=9.64cm}}
\end{picture}
\end{center}
\caption{Scaling property of the linear specific heat coefficient, for
$n=1.2$ and various $t$. Squares: $t=0.2$~eV. Circles: $t=0.3$~eV, range to
12~K scaled by $c=4.75$ in Eq.~(\ref{eqsca}). Full line: $t=0.8$~eV, range to
100~K, $c=127.4$. Dashed line: $t=1.0$~eV, range to 100~K, $c=232.7$.
\label{gamma}}
\end{figure}
Indications to the same effect can be obtained studying the bulk
effective mass, or, technically, the linear specific heat coefficient
$\gamma=c_V/T$. First, in non-interacting band electrons, $\gamma(T)$ obeys a
simple scaling relationship, when the overlap $t$ is changed:
\begin{equation}
\gamma_{t'}(T)=c\gamma_t(T/c).
\label{eqsca}
\end{equation}
This is a `law of corresponding states' for the Fermi gas: the system with an
overlap reduced to $t'<t$ is the same as the `old' system at a lower
temperature and higher effective mass ($c>1$). Figure \ref{gamma} shows that
this is obeyed to very high precision for states in the in-gap band. It
should be noted, however, that the renormalizations involved are
quantitatively much larger than in the non-interacting case. For instance,
changing $t$ from 1~eV to 0.2~eV involves a factor $c$ of only about 10 for
non-interacting electrons, while the corresponding factor in figure
\ref{gamma} is over 200.

By contrast, in the lower Hubbard band, the scaling (\ref{eqsca}) is not
obeyed. This is consistent with the behavior of the interaction contribution
to the entropy: the system being a liquid there, interactions spoil the
scaling which depends only on the kinetic parameter $t$.

A further characteristic of a first-order transition is that the effective
mass does not diverge. The same behavior as in figure \ref{gamma} is also
observed for fillings in the immediate vicinity of the transition. The bulk
mass passes through a maximum and saturates, even though for the small value
$t=0.2$~eV shown in the figure, this only happens below $0.5$~K. The two
minima in the free energy, corresponding to the lower Hubbard band and in-gap
band, exchange place without the fluctuations around either diverging.

\section{Discussion}

The quantum transition described here differs from usual model descriptions
of the Mott-Hubbard transition, in that it occurs at an incommensurate
filling. This is disconcerting, since the classical `counting' prediction for
the transition point is upheld by particle-hole symmetry. However, this
symmetry is broken by forbidding fluctuations onto a doubly occupied site. In
the one-band model, this results in the disappearance of phase space
precisely at half-filling. In the three-band model, there is still phase
space associated with the oxygen sites, so there are quantum fluctuations
left, even if the doubly occupied site is treated classically. This is the
regime of the present model.

The quantum fluctuations are due to projected hopping, which plays the role
of an interaction. It creates the in-gap band, which explains why it
corresponds to a `gas' phase: this is where hopping has won! Once
particle-hole symmetry is broken, the criterion for the transition becomes
quantitative, as expressed by equation (\ref{crit}). Such a situation is
generic for a phase transition, whose position is usually determined by a
competition of energy scales, not arguments of symmetry. Magnetic
interactions, for instance, also cause a transition in some incommensurate
range, around half-filling. In their absence, no long-range order is expected
in the preset model, although a significant tendency to order may be inferred
from the interaction entropy in the lower Hubbard band, as the transition is
approached~\cite{sunko}.

Charge correlations prefer the less ordered `gas' phase when space becomes
crowded, and so their effect is opposite to that of magnetic interactions.
The in-gap states are called a gas because, as discussed in section
\ref{trans}, they are characterized by an absence of residual interactions.
As opposed to these qualitative considerations, their quantitative parameters
may indicate states very close to localization, as shown in
figure~\ref{gamma}. An interpretation of the gas in terms of the underlying
fermions is not possible directly within the model, because it uses a trick
to count the fermion phase space correctly: the $k$-space states which
diagonalize the grand potential are based on the translational invariance of
the ensemble as a who\-le, not of an individual member of it. The states in a
\emph{given} sample could be different from a Fermi liquid. The `gas'
probably means that annealment by quantum fluctuations has moved the
forbidden sites out of the way, for hopping to occur down (quantum)
percolation channels, which extend across the system, giving rise to some
effective band-width. This gives an appealing picture, how mean free paths
can be shorter than the lattice spacing: at low $t/\Delta_{pd}$, few of these
channels remain open at any finite temperature, because hopping is then
inefficient against the even larger entropy associated with complete site
disorder. Simulations should help confirm this interpretation, but very steep
increases of the entropy with temperature have been obtained~\cite{sunko} in
the model at $t/\Delta_{pd}\sim 1/10$.

Finally, one may wonder how much of this picture would remain, if both kinds
of electrons were allowed to hop. It is easy to imagine some kind of mutual
compromise along the lines sketched above, but the physical question is,
would such a network survive on time scales much longer than those associated
with the traversal of a single electron across the crystal. If so, this would
give a picture of conduction in real space, in which a given electron moves
as a single particle, but all electrons of one spin present themselves as a
quasi-static collective (the network) to those of the other spin. The present
model is essentially a realization `by hand' of this intriguing symmetry
breaking, first proposed by Gutzwiller for the Hubbard model: that one kind
of spin sees the other `as if occupying a band of width zero'~\cite{gutz}.
Interestingly enough, the model construction implies a direct experimental
consequence of this assumption. Namely, the chemical potential of these
`other' spins, the $\nu$ in equation (\ref{fact}), is independent of doping
throughout the in-gap band (because they are dispersionless), so the
dispersion $\varepsilon(\nu)$ should not vary with doping either.
Experimentally, positions of dispersive peaks along typical cuts in the
Brillouin zone vary by a rough, but still unexpectedly uniform, 0.2~eV in a
wide class of materials~\cite{ybco,bscco}, from optimally doped to
insulators~\cite{shen}, but again, this changes for overdoping~\cite{shen}.

To conclude, a model quantum phase transition has been described, caused by
hopping fluctuations in the presence of a classical on-site repulsion. It is
analogous to a liquid-gas transition, and is not associated with a divergence
of the effective mass, even though it may appear otherwise for all but the
lowest temperatures.

\section{Acknowledgements}

Conversations with S. Bari\v si\'c and one with A. Georges are gratefully
acknowledged. This work was supported by the Croatian Government under Project
$119\,204$.

\end{document}